\documentclass[nofootinbib,twocolumn]{an}
\usepackage{amsmath}
\usepackage{graphicx}
\usepackage{times}

\begin{document}

\title{Supermassive black hole mergers as dual sources for electromagnetic
flares in the jet emission and gravitational waves}
\author{ M.\, T\'{a}pai\inst{1}\fnmsep\thanks{Corresponding author:
  \email{tapai@titan.physx.u-szeged.hu}} \and L.\,\'{A}.\, Gergely \inst{1,2} \and Z.\, Keresztes\inst{1} \and P.\, J.\, Wiita%
\inst{3} \and Gopal-Krishna\inst{4} \and P.\, L.\, Biermann\inst{5,6,7,8,9}}

\authorrunning{T\'{a}pai et al.} 
\titlerunning{SMBH mergers as dual sources for EM flares in the jet emission and GWs}

\institute{Departments\ of Theoretical and Experimental Physics, University\ of
Szeged, Hungary
\and
Department of Physics, Tokyo University of Science, Shinjuku-ku, Tokyo, Japan
\and
Department\ of Physics, The College of New Jersey, Ewing, NJ, USA
\and
National Centre for Radio Astrophysics, TIFR, Pune, India
\and
Max Planck Institute for Radioastronomy, Bonn, Germany
\and
Department\ of Physics \& Astronomy, University\ of Bonn, Germany 
\and
Department\ of Physics \& Astronomy, University\ of Alabama,
Tuscaloosa, AL, USA
\and
Department\ of Physics, University\ of Alabama at Huntsville, AL, USA
\and
FZ Karlsruhe and Physics Department, University of\ Karlsruhe, Germany}


\abstract{
We present a new type of observation relating the gravitational wave emission of 
supermassive black hole mergers to their electromagnetic counterparts.  This dual emission involves
variability of a relativistic jet arising from the spin-orbit precession of the 
supermassive black hole binary at its base. }
\maketitle
\section{Introduction}

The large number of parameters of the gravitational wave (GW) sources
induces degeneracies obstructing their identification from GW observations
alone. Coincident detections of electromagnetic (EM) and GW signatures from
supermassive black hole (SMBH) coalescence would significantly improve this
situation.

EM signatures from SMBH binaries could arise through  interaction with
the surrounding matter. An early attempt\ to estimate the EM counterparts of
Advanced LIGO and Virgo GW sources (neutron star-neutron star or neutron
star-black hole binary) by Sylvestre (2003)\nocite{Sylvestre} reviewes three different
mechanisms: magnetospheric interactions, radioactive decay of ejected
material, and relativistic blast waves.

Some of the theoretical studies model the merging BH binary as immersed in a
gas cloud or surrounded by a circumbinary disk (Armitage \& Natarajan 2002,\nocite{circumbinary} 
2005;\nocite{armitnat} Cuadra et al. 2009;\nocite{CAAB} Dotti et al. 2007;\nocite{DCHM} Escala et al. 2005;\nocite{Escala} Roedig et
al. 2011)\nocite{roedig}.  Other models involve interaction with a nuclear star cluster (Khan, Just, \&
Merritt 2011;\nocite{Merritt} Preto et al. 2011;\nocite{Preto} Zier \& Biermann 2001,\nocite{ZierB1} 2002)\nocite{ZierB2}. The hardening
of the binary by angular momentum and energy exchange with stars was also
investigated (Berczik et al. 1996\nocite{BMSB}; Perets \& Alexander 2008\nocite{perets}; Quinlan 1996\nocite{hardening};
Sesana, Haardt \& Madau 2007\nocite{Sesana}). Observational evidence is restricted
for such disks, as direct observations of merging SMBH
systems are very sparse; the best claim (Rodriguez et al. 2006\nocite{Rodriguez}) has been supported by
Morganti, Emonts and Oosterloo (2009\nocite{Morganti}).%

Krolik (2010\nocite{Krolik}) has shown that the duration of the EM signal is proportional
to the mass of the gas near a pair of merging BHs, and rather longer than
the merger event. Both the realignment of the gas orbit to the spin
orientation of the newly formed BH and the orbital adjustment to the mass
lost by GW augment the EM radiation over longer timescales.

Milosavljevic and Phinney (2005\nocite{MPhinney}) have studied how the binary hollows out the
surrounding gas and shrinks slowly compared to the viscous timescale of a
circumbinary disk. The truncation of the inner gas disk at a radius where
gravitational torque from the binary equals the viscous torque leads to a
diminished accretion onto the black holes. The viscous evolution of the
hollow disk will be visible in X-rays. According to a Newtonian calculation
of the EM afterglow (Shapiro 2010\nocite{Shapiro}) both the temporal increase in the total
EM flux and the hardening of the spectrum will confirm the interpretation of
a GW interferometer signal as arising from the merger of a binary BH.

Arguments supporting the occurrence of the afterglow much sooner after the
merger than previously estimated have been presented by Tanaka and Menou
(2010\nocite{TanMen}). The birth of a quasar as triggered by the merger of the SMBH host
galaxies is then preceded by this afterglow, and it is proposed that all-sky
soft X-ray surveys could identify them (Tanaka, Haiman \& Menou 2010\nocite{THM}).
Multiple EM flares from tidally disrupted stars could also follow the SMBH
binary merger (Stone \& Loeb 2011\nocite{StoneLoeb}).

The extent to which SMBH mergers in quasars could be detected through GW
emission has been discussed by Kocsis et al. (2006\nocite{Kocsis06}). The size and orientation of the three-dimensional error ellipse in solid angle and redshift within which a LISA (Laser Interferometer Space Antenna) (Arun et al. 2009\nocite{LISA}) event could be localized using the GW signatures was given. Assuming that BH mergers are accompanied by gas accretion leading to Eddington-limited quasar activity, the number of quasars in a typical LISA error volume gave at low redshifts a single near-Eddington quasar at z=1. For rapidly spinning SMBHs this result may be extended up to z=3. The possibility that a
dedicated optical or X-ray survey could identify coalescing SMBH binaries
statistically, as a population of periodically variable quasars has been
discussed (Haiman, Kocsis \& Menou 2009\nocite{HKM}).

Dotti et al.\ (2006\nocite{DSSCH}) have considered the mutually exclusive possibilities of
detecting either an EM precursor, during the last year of a GW-driven
inspiral, or an afterglow within a few years after coalescence. The
precursors correspond to on-off states of accretion on to a primary SMBH
heavier than $\sim 10^{7}$M$_{\odot }$ (a bright X-ray source decaying), and
they are associated with galaxies containing ongoing starbursts. Lighter
binaries, by contrast, exhibit an off-on accretion flow rising in $<20$
years, leading to an EM afterglow. The possible recoil of the merged SMBH
could produce strong shocks, resulting in an afterglow with characteristic
photon energy increasing in time from the UV to the soft X-ray range,
between one month and a year after the merger (Lippai, Frei \& Haiman 2008\nocite{Lippai08}),
provided the Schnittman and Buonanno (2007\nocite{SBuonanno}) empirical formula for the
dependence of the kick velocity on spins and mass ratios is adopted. The
possibility of detecting prompt EM counterparts was discussed by Kocsis,
Haiman and Menou (2008\nocite{KHM}), both for transient signals and by cross-correlating
the period of any variable EM signal with the quasi-periodic gravitational
waveform over 10--1000 cycles. The effect of EM counterparts to GWs from a
binary BH system on plasmas and EM fields in their vicinity was also
considered (Palenzuela, Lehner \& Yoshida 2010\nocite{Palen}). The binary's dynamics
induces time dependence in the form of a variability in electromagnetically
induced emissions and an enhancement of EM fields in the final stages of the
merger.

Using a general relativistic, hydrodynamical study of the late inspiral
phase and merger of equal-mass, spinning SMBH binaries immersed in hot gas
flows, Bode et al.\ (2010\nocite{Bode}) have argued that variable EM signatures
correlated with GW emission can arise due to shocks, accretion, relativistic
beaming, and Doppler boosting modulated by the binary orbital motion. The
effect is largest for binary systems with the individual BH spins aligned
with the orbital angular momentum. The frequency of the EM oscillations and
GWs was found to be equal and the variations in luminosity to be within a
factor of 2. The most massive binaries detectable in the LISA band may be
identified in EM searches out to $z=1$, for a high enough gas density.

A binary moving in a uniform magnetic field anchored to a circumbinary disk
was considered by M\"{o}sta et al. (2010\nocite{Mosta}) for configurations where the spins
were either aligned or anti-aligned with the orbital angular momentum. The
emitted EM waves mimic the phase of the GWs but have quite small amplitudes
and peak at frequencies unaccessible to radio astronomical observations. In
particular, the energy emission in EM waves was shown to be 13 orders of
magnitude smaller than in GW and the corresponding luminosity is also much
smaller than the accretion luminosity for systems accreting close to the
Eddington rate. Nonetheless it was conjectured, that with a small and stable
accretion rate of the circumbinary disk over the timescale of the final
inspiral, the EM emission will alter the accretion rate through magnetic
torques; hence it may be observable indirectly.

In another mechanism the GWs themselves induce EM waves propagating away
through the ambient gas, and shear which is eventually dissipated as heat
(Kocsis \& Loeb 2008\nocite{KL08}).

The symbiotic systems of black holes, accretion disks and magnetospheres
however include another important element, not yet considered. This is the
energetic jet a SMBH often emits in the direction of its spin. The SMBH spin and the jet spectrum correlate (Kun et al. in prep.\nocite{kemma}). When a SMBH lying
at the base of the jet is moved around by the inspiral of a smaller black
hole, then a violent EM wave travels along the magnetic field structure of
the jet and the geometry of the jet at its base is also distorted. 

The purpose of this paper is to investigate in a simple model how the
observations on the periodicity in a jet at the base of which there is a
SMBH binary could result in complementary information for GW detection by
the LISA space mission.

\section{Flares in the jet spectrum due to spin-orbit precession}

Most of the jets will have an orientation quite far from the line of sight;
however,  many jets pointing close to us are detected by radio techniques.
This detection is enhanced by a strong selection effect:  relativistic boosting of
the emission from the jet is so powerful, that half of all radio sources
detected at 5 GHz are relativistic jets pointed nearly at us (Chini et al. 1988\nocite{CSKKQSW};
Eckart et al. 1986\nocite{eckart01}, 1987\nocite{eckart02}, 1989\nocite{eckart03}; Gregorini et al. 1984\nocite{Gregorini1984}; K\"{u}hr et al.
1981a\nocite{jetselection}, 1981b\nocite{khur}).

If for some reason a jet shows a precessional evolution, and it is also
relativistic, when it comes close to our line of sight, it will produce
significant variability at all wavelengths, mostly detectable in the radio,
hard X-ray and gamma-ray spectrum. Such a precession could originate from
the spin-orbit interaction in a SMBH binary lying at the base of the jet.
Due to this interaction the spins undergo a precessional motion about the
orbital angular momentum $\mathbf{L}$, each spin sweeping over a cone.
Gergely and Biermann (2009\nocite{spinflip1}) explored in detail the consequences of the
simultaneous spin-orbit precession and GW backreaction. For the typical mass
ratio range of $1:3$ to $1:30$ it was shown that the second spin can be
neglected as $S_{2}/S_{1}=\nu ^{2}\chi _{2}/\chi _{1}$ (where $\chi
_{1,2}=cS_{1,2}/ Gm^{2}_{1,2} $ are the dimensionless spin
parameters, $G$ is the gravitational constant, $c$ the speed of light and $%
\nu =m_{2}/m_{1}<1$ the mass ratio). Thus the precession of the dominant
spin $\mathbf{S}_{\mathbf{1}}$ can be regarded to occur about the total
angular momentum $\mathbf{J}=\mathbf{L}+\mathbf{S}_{\mathbf{1}}$. For this
mass ratio range a reorientation of $\mathbf{S}_{\mathbf{1}}$ occurs during
the inspiral. In the process the cone swept by the jet will change its opening
angle. As at the end of the inspiral the direction of the dominant spin
becomes closely aligned to the total angular momentum, the opening of the
cone becomes quite narrow. The spin-flip can be visualized as the narrowing
in time of the precession cone until $\mathbf{S}_{\mathbf{1}}$ becomes
quasi-aligned to $\mathbf{J}$. Therefore, any jet variability detected due
to this mechanism will be a transient phenomenon.

\begin{figure}[tbp]
\begin{center}
\includegraphics[width=6.5cm]{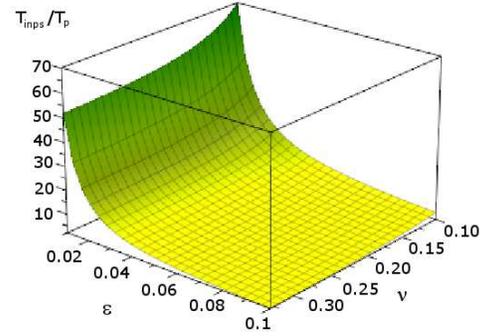}
\end{center}
\caption{The ratio $T_{\mathrm{insp}}/T_{\mathrm{p}}$ of the GW and precession time-scales,
represented for the mass ratio range $\nu \in [0.1,0.3]$ and the PN parameter range 
$\varepsilon \in [0.02,0.1]$. }
\label{tGWtp}
\end{figure}

The period $T_{p}=2\pi \Omega _{p}^{-1}$ of the precession (where $\Omega
_{p}$ is the precessional angular velocity) gives the precession timescale,
approximated as (Gergely \& Biermann 2009\nocite{spinflip1}) 
\begin{equation}
\frac{1}{T_{\mathrm{p}}}\approx \frac{c^{3}\eta }{\pi Gm}\varepsilon ^{5/2}\frac{J}{L}%
\approx \frac{c^{3}\eta }{\pi Gm}\varepsilon ^{5/2}\sqrt{1+\left( \frac{S_{1}%
}{L}\right) ^{2}}~,  \label{cond1}
\end{equation}%
where $m=m_{1}+m_{2}$ is the total mass, $\eta =m_{1}m_{2}/m^{2}=\nu /\left(
1+\nu \right) ^{2}$ the symmetric mass ratio, $\varepsilon =Gm/c^{2}r$ the
post-Newtonian (PN) parameter and $r$ the orbital separation. For the last
approximation $\mathbf{S}_{\mathbf{1}}$ and $\mathbf{L}$ were taken as
perpendicular, in order to be able to further employ 
\begin{equation}
\frac{S_{1}}{L}=\varepsilon ^{1/2}\nu ^{-1}\chi _{1}~.  \label{S1L}
\end{equation}%
The typical time-scale of the inspiral is%
\begin{equation}
\frac{1}{T_{\mathrm{insp}}}=-\frac{\dot{L}}{L}=\frac{32c^{3}}{5Gm}\eta \varepsilon
^{4}~,  \label{cond2}
\end{equation}%
hence the ratio of these time-scales becomes%
\begin{equation}
\frac{T_{\mathrm{insp}}}{T_{\mathrm{p}}}=\frac{5}{32\pi }\varepsilon ^{-3/2}\sqrt{1+\left(
\varepsilon ^{1/2}\nu ^{-1}\chi _{1}\right) ^{2}}~.
\end{equation}%
This ratio is smallest in the last stages of the inspiral, represented
on Fig. \ref{tGWtp}, indicating that the time-scale under which GWs cause
significant changes in the orbit is typically two orders of magnitude larger
than the precession time-scale. In this stage an observed variability in the
jet of the order of a day, if due to rapid precession would imply few remaining months for the evolution
of the binary until the merger and related GW emission. 

\section{Orbital characteristics of the binary from observations}

For simplicity we assume that the dominant SMBH is maximally spinning ($\chi
_{1}=1$). Let the angles between $\mathbf{S}_{\mathbf{1}}$ and $\mathbf{J}$
be $\beta $, while between $\mathbf{S}_{\mathbf{1}}$ and $\mathbf{L}$ be $%
\kappa $ (which, to leading order GW radiation stays constant, Gergely, Perj%
\'{e}s \& Vas\'{u}th 1998\nocite{GPV3}). Additionally, the law of sines and Eq. (\ref%
{S1L}) allow us to express $\kappa $ as 
\begin{equation}
\kappa =\beta +\arcsin \left[ \varepsilon ^{1/2}\nu ^{-1}\sin \beta \right]
~.
\end{equation}

\begin{table}[tbp]
\caption{$T_{\Delta \protect\beta }$ and $T_{\mathrm{p}}$ calculated for $\protect\nu %
=0.1$\textbf{, }$\protect\varepsilon _{\Delta \protect\beta }=0.01$ and $%
m=10^{6}~M_{\odot }$ as function of the angle $\protect\beta $. The value of 
$\protect\kappa $ is also indicated.$\ $}
\begin{tabular}{cccc}
\hline
$\beta ~[^{\circ }]$ & $\kappa ~[^{\circ }]$ & $T_{p}~[\mathrm{days}]$ & $T_{\Delta
\beta }~[\mathrm{days}]$ \\ \hline
$20$ & $40$ & $116$ & $1041$ \\ 
$25$ & $50$ & $120$ & $812$ \\ 
$30$ & $60$ & $126$ & $656$ \\ 
$35$ & $70$ & $133$ & $541$ \\ 
$40$ & $80$ & $142$ & $451$ \\ \hline
\end{tabular}%
\label{table03}
\end{table}

We identify the period of the flares in the jet with the $T_{\mathrm{p}}$. A more
accurate estimate combines the next-to-the-last expression (\ref{cond1}) and
the law of sines as 
\begin{equation}
T_{\mathrm{p}}\left( \varepsilon ,\nu ,m,\beta \right) =\frac{\pi Gm\left( 1+\nu
\right) ^{2}}{c^{3}\varepsilon ^{5/2}\nu }\frac{\sin \beta }{\sin \kappa }~.
\end{equation}%
As the frequency of the gravitational wave to leading order is%
\begin{equation}
f_{\mathrm{GW}}=\frac{c^{3}}{\pi Gm}\varepsilon ^{3/2}~,  \label{fGW}
\end{equation}%
the jet variability period can also be expressed as 
\begin{equation}
T_{\mathrm{p}}\left( \varepsilon ,\nu ,f_{\mathrm{GW}},\beta \right) =\frac{\left( 1+\nu
\right) ^{2}}{\varepsilon \nu }\frac{\sin \beta }{\sin \kappa }f_{\mathrm{GW}}^{-1}~.
\end{equation}

The GW frequency $f_{\mathrm{GW}}$ and PN parameter $\varepsilon $ determine the time 
$T_{\Delta \beta }$ characterizing how long the jet variability is observed.
Gergely and Biermann (2009\nocite{spinflip1}) have derived the evolution of $\kappa -\beta $
under radiation reaction (their Eq. (36)). Employing the $\sin $ theorem we
find%
\begin{equation}
\dot{\beta}\approx -\frac{32c^{3}}{5Gm}\frac{\varepsilon ^{9/2}\sin
^{2}\beta }{\left( 1+\nu \right) ^{2}\sin \kappa }~.
\end{equation}%
We approximate with $\Delta \beta =-\int_{0}^{T_{\Delta \beta }}\dot{\beta}%
dt\approx -\dot{\beta}_{0}T_{\Delta \beta }$ the angle under which the
variability in the jet spectrum is still detectable (the variability is
observable in some range $( \beta _{0}-\Delta \beta /2, \beta
_{0}+\Delta \beta /2 ) $). Hence $T_{\Delta \beta }$ becomes entirely
determined by the set $\varepsilon ,\nu ,m,\beta ,\Delta \beta $ (or, by
employing Eq. (\ref{fGW}), by $\varepsilon ,\nu ,f_{\mathrm{GW}},\beta ,\Delta \beta $
): 
\begin{equation}
T_{\Delta \beta }=\frac{5\Delta \beta }{32\pi }\frac{\left( 1+\nu \right)
^{2}\sin \kappa }{\varepsilon ^{3}\sin ^{2}\beta }f_{\mathrm{GW}}^{-1}~.
\end{equation}%
The ratio $T_{\Delta \beta }/T_{p}$ has a simpler parameter dependence%
\begin{equation}
\frac{T_{\Delta \beta }}{T_{p}}\left( \varepsilon ,\nu ,\beta ;\Delta \beta
\right) =\frac{5\Delta \beta }{32\pi }\frac{\nu \sin ^{2}\kappa }{%
\varepsilon ^{2}\sin ^{3}\beta }~.
\end{equation}

For a given mass ratio $\nu $, given inclination $\beta $ and a given
observational sensitivity $\Delta \beta $ the set of observables $T_{\mathrm{p}}$ and 
$T_{\Delta \beta }/T_{\mathrm{p}}$ will determine the PN\ parameter $\varepsilon
_{\Delta \beta }$ and the associated GW frequency $f_{\mathrm{GW}}$ (or,
equivalently, the total mass).

For illustration, in Table \ref{table03} we give $T_{\Delta \beta }$\ and $%
T_{\mathrm{p}}$\ for $\nu =0.1$, $\varepsilon _{\Delta \beta }=0.01$ and $%
m=10^{6}~M_{\odot }$\ for various\ values of $\beta $, and $\Delta \beta = 1^{\circ}$. 

\section{Concluding Remarks}

SMBH encounters typically do not have equal masses, nor do they usually have
extreme mass ratios, implying that a spin flip very likely happens during
the inspiral. Such spin-flips provide a mechanism to form X-shaped RGs via a
rapid reorientation of the jet direction (Gopal-Krishna et al. 2012\nocite{XRG}; Mezcua
et al. 2011\nocite{Mezcua1}, 2012\nocite{Mezcua2}). The spin-flip can be visualized as the narrowing in time
of the precession cone until $\mathbf{S}_{\mathbf{1}}$ becomes quasi-aligned
to $\mathbf{J}$. A jet swerving into the line of sight would immediately
constitute an extreme brightening of a flat spectrum radio source, and
precession of such a source on time scale of a few days would produce very
strong flaring. All extremely variable AGN are flat spectrum radio sources
near 5 GHz (Eckart et al. 1986\nocite{eckart01}; Gregorini et al. 1984\nocite{Gregorini1984}). Therefore such a
source would be immediately recognizable, and some of the most extreme
flaring sources among the known flat spectrum radio sources are clear
candidates to be very close to a merger.

Any jet variability detected due to this mechanism will be a transient
phenomenon, as the cone continues to narrow. Assuming that such strong
variability in the jets is detected as a transient phenomena, there will be
two timescales given by the jet observations: the period of the variability
(we identify this with the spin-orbit precessional period) and the time
 the transient phenomenon lasts, as well as an angle $\Delta \beta $
determined by the sensitivity of the observations. We have determined the
dependence of these observables on the mass ratio $\nu $, PN parameter $%
\varepsilon $, jet inclination $\kappa $ vs. the orbital plane and GW
frequency $f_{\mathrm{GW}}$, at the time of observation. These additional jet
observables greatly contribute to reduce the known degeneracies in the
parameter space by pure GW detection.

\acknowledgements
L\'{A}G was supported by the EU grants T\'{A}MOP-4.2.2.A-11/1/KONV-2012-0060, T\'{A}MOP-4.2.2.C-11/1/KONV-2012-0010 and the Japan Society for the Promotion of Science.

\end{document}